\documentclass[twocolumn,pra,amsmath,amssymb]{revtex4}
\usepackage{graphicx}
\usepackage{dcolumn}
\usepackage{bm}

\newcommand{\eg}{{\it{e.g.}}}
\newcommand{\ie}{{\it{i.e.}}}

\newcommand{\refeq}[1]{Eq.~(\ref{#1})}
\newcommand{\Tr}{\mathrm{ Tr }}
\def\one{\leavevmode\hbox{\small1\normalsize\kern-.33em1}}

\begin{document}
\title{Local temperature in quantum thermal states}

\author{Artur Garc\'\i a-Saez$^1$, Alessandro Ferraro$^1$, and Antonio Ac\'\i n$^{1,2}$}
\affiliation{$^1$ ICFO--Institut de Ci\`{e}ncies Fot\`{o}niques,
Mediterranean Technology Park, 08860 Castelldefels (Barcelona),
Spain\\
$^2$ ICREA-Instituci\'o Catalana
de Recerca i Estudis Avan\c cats, 08010
Barcelona, Spain
}

\begin{abstract}
We consider blocks of quantum spins in a chain at thermal equilibrium, focusing on their properties from a thermodynamical perspective. In a classical system the temperature behaves as an intensive magnitude, above a certain block size, regardless the actual value of the temperature itself. However, a deviation from this behavior is expected in quantum systems. In particular, we see that under some conditions the description of the blocks as thermal states with the same global temperature as the whole chain fails. We analyze this issue by employing the quantum fidelity as a figure of merit, singling out in detail the departure from the classical behavior. As it may be expected, we see that quantum features are more prominent at low temperatures and are affected by the presence of zero-temperature quantum phase transitions. Interestingly, we show that the blocks can be considered indeed as thermal states with a high fidelity, provided an effective local temperature is properly identified. Such a result may originate from typical properties of reduced sub-systems of energy-constrained Hilbert spaces. Finally, the relation between local and global temperature is analyzed as a function of the size of the blocks and the system parameters.
\end{abstract}

\date{\today}
\maketitle

\section{Introduction}

Since the early days of quantum theory, there has been a considerable
effort to formulate the principles of thermodynamics from a quantum
perspective \cite{MahlerBook}. Recently, this approach has allowed to
discover novel features of the nature of many-body systems, see for
instance Refs.  \cite{tasaki,malPRL,malPRE,psw,lebo}. In particular,
careful analysis of the application of thermodynamics concepts to
microscopic systems have shown the appearance of purely quantum
features. In such systems, thermodynamic magnitudes may indeed loose
their classical properties, giving rise to peculiar dependencies on
some system parameters, such as size or total energy. A striking
manifestation of this is given by the temperature, a magnitude
considered to be intensive in classical thermodynamics. In a quantum
scenario the temperature may not be well defined, resulting in the
fact that subparts of thermal states may no longer be described as
thermal states with the same global temperature as the whole system.
More precisely, recent studies suggest that the intensive nature
of the temperature may be lost not only in dependence of the size of
the system subparts, as in a classical scenario, but also changing the
temperature of the global system \cite{malPRL,malPRE}. It is then a
relevant question to understand under which conditions the concept of
temperature offers a correct description of sub-parts of quantum
thermal states.


Although the main motivation of this work comes from a fundamental point of view, studying the limits of validity of thermodynamics concepts may be relevant from a practical perspective. In fact, recent experimental progress in nano-sciences allows to access thermodynamical quantities, like temperature, at scales in which deviations from classical thermodynamics may become relevant \cite{nanotherm}. For example, in Refs. \cite{mahlEPL,HartCP} it has been pointed out that the breakdown of the concept of temperature might have consequences on thermometry and might be observed in experiments with spin chain compounds.

In Refs.~\cite{malPRL,malPRE} a set of conditions were established in
order to assure that a large thermal system can be approximated by a
set of {\em factorized} blocks, each of them described in turn by a
thermal state at the same temperature as the global system. Clearly,
when such an approximation is valid, local measurements performed on a
block provide results compatible with the global temperature. In other
words, we do meet a situation in which temperature is intensive.
However, local measurements on a single block do not provide any
information about the correlations with the other blocks, which are
simply disregarded (traced out) in the measuring process. Therefore, the approximation in Refs.~\cite{malPRL,malPRE} may be
further relaxed still retrieving situations compatible with the
concept of intensive temperature. As a matter of fact, one may only
require that the block actually measured should be in a thermal state
at the same temperature as the global one, not necessarily factorized with
the rest of the system. This is the approach adopted in this work.

The aim of the present work is to consider these issues exploiting
ideas and tools recently emerged from quantum information science.
This approach has the advantage of giving a thorough description of
the quantum systems under consideration, allowing in turn to
characterize in detail the departure from the classical behavior.  We
consider a chain composed by $n$ spins (where $n$ can be taken in the
macroscopic limit) at temperature $T$ and focus on the thermodynamical
properties of a block composed by $m$ spins ($m<n$) --- \ie, the
reduced state obtained after tracing out $n-m$ spins.
In a standard thermodynamic setting $m$ and $n$ are taken large enough such that the interactions between the block and the rest of the system may be disregarded. As a consequence, the block can be well described by a thermal state at the same temperature as the global one. However, such a picture may brake down for blocks of small size and strong interactions. As said, whereas for classical systems this breakdown has no dependence on the temperature, in quantum systems a temperature dependence arises \cite{malPRL}. This is the scenario that we will consider here.
To face this problem we use the quantum fidelity as a figure of merit.  The latter
quantifies the amount of statistical distinguishability between two
quantum states and its properties have been widely studied in quantum
information science \cite{NC}.  Recently, the fidelity has been used
both as an indicator of phase transitions in spins systems
\cite{zan1,zan2}, and to study the emergence of thermal states in
contiguous blocks of some spin-system ground states~\cite{olav}.  As
shown in these works, the fidelity has revealed to be a particularly
sensitive figure of merit in this framework. Here, we use it to check
whether the reduced states of the spin chain under consideration can
be well approximated by an $m$-particle thermal state.
The sensitivity of this approach will be assessed by investigating regions
in the phase space near the zero-temperature critical points.


We also relate our investigations to recent findings in the foundation of
statistical mechanics -- again inspired by quantum information
concepts. In Refs.~\cite{psw,lebo} the authors address typical
properties of the reduced states obtained by tracing out a huge amount
of degrees of freedom from a constrained pure system. Building on previous
results on properties of quantum states in large dimensional systems~\cite{HLW06},
it was shown in Ref.~\cite{psw} that the reduced state of a big system (including
environment) satisfying an operator constraint is basically the same for almost
any pure state of the system. In the particular case in which (i) the operator
constraint is related to the energy of the whole system and (ii) the interaction
between system and environment is small, this typical state can be shown to
correspond to a canonical thermal state~\cite{psw,lebo}, as already pointed out
in the early days of quantum mechanics \cite{schro}.

Motivated by these results we study the reduced states of the chain from a
thermodynamical perspective, describing it with only a few physical magnitudes,
an effective temperature in our case. In other words, we check whether the reduced
states of thermal systems maintain some sort of canonical typicality. Recall that the
results in Refs.~\cite{psw,lebo} show that a canonical thermal state for the reduced
system is obtained when the interaction energy between the considered parts is
negligible. As said, here we will consider settings in which such condition is not
fulfilled, hence the validity of these typicality results is by no means guaranteed.
Specifically, we proceed by identifying the thermal state of $m$
particles that, subject to the local interaction inherited by the
whole Hamiltonian, is closer to the actual reduced state of the whole
chain. We thus define an effective {\it local} temperature (the only
free parameter to adjust) that can be compared with the {\it global}
temperature of the whole system. Depending on the parameters of the
Hamiltonian and the subsystem size we find situations where the local
temperature is no longer equal to that of the global system, in
accordance with the above mentioned analysis. However, the description
of the reduced states as thermal states is an extremely good
approximation, as shown by the corresponding fidelity.
This is remarkable since, as said, we are not in the standard
conditions considered in Refs.~\cite{psw,lebo}. Our results then suggest
that some form of canonical typicality may still be present, even if the
interaction between system and environment is not negligible. In particular,
we find that below some threshold temperature the local temperature may
become higher than the global one, and that the reduced states can have some
finite temperature even when the global system is in the ground state.
Furthermore, as one may expect, the local and global
temperatures tend to coincide increasing the size of the subsystem.


The structure of the paper is the following. After introducing the
systems under consideration, we calculate in Section
\ref{s:intensive_T} the fidelity between two-particle blocks and the
corresponding thermal state. This allows to study the conditions under
which the temperature ceases to be an intensive magnitude.  Then, in
Section \ref{local}, we derive the local effective temperature of the
two-particle blocks, as well as the fidelity of this description. The
dependence of the local temperature on the size of the blocks is
studied in Section \ref{larger}, where blocks consisting of more than
two spins are studied with the help of the Matrix Product States
formalism.  Finally, we summarize our results in Section \ref{esco}
and point out some conclusive remarks.

\begin{figure}
\includegraphics[width=8cm]{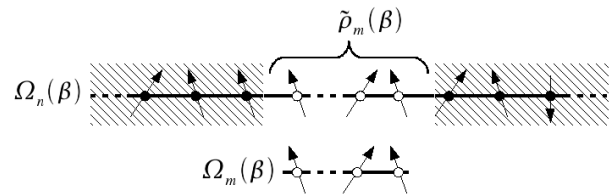}
\caption{\label{fig:diag} (Upper panel) Reduced $m$-particle state
  $\tilde{\rho}_m(\beta)$ of a chain composed by $n$ particles in the
  thermal state $\Omega_n(\beta)$ at temperature $T=1/\beta$. The
  state $\Omega_n(\beta)$ is built from an $n$-particle Hamiltonian
  $\hat{H}_n$. (Lower panel) Thermal state $\Omega_m(\beta)$ of $m$
  particles for the same local interaction $\hat{H}_m$ but referred to
  $m$ particles.}
\end{figure}

\section{Intensive temperature analysis}
\label{s:intensive_T}


We start by assessing the applicability of the concept of intensive
temperature. As said, we consider a large spin system at temperature
$T=1/\beta$ and focus on a sub-block of spins. We then check whether
the latter can be well approximated by the thermal state at
temperature $T$ given by the interaction inherited from the whole
system Hamiltonian.
By analyzing the relation between the reduced states of thermal systems
and thermal states themselves, we assess the validity of such estimation.


Consider the setting depicted in Fig.~\ref{fig:diag}. On the one hand,
we construct the canonical state of $n$ particles $\Omega_n(\beta)=
\exp{(-\beta \hat{H}_n)}/Z_n$, where $\hat{H}_n$ is a local
interacting Hamiltonian and $Z_n$ is the partition function. In order
to check the thermal properties of the subparts of this state we trace
out a part of it, obtaining the state of $m$ particles,
$\tilde{\rho}_m(\beta)= Tr_{n-m}[\Omega_n(\beta)]$ (throughout this
work we will focus on systems composed by a number of particles $n$
much larger than $m$). On the other hand, we directly construct the
canonical state of $m$ particles $\Omega_m(\beta)= \exp{(-\beta
  \hat{H}_m)}/Z_m$.  We then compare these two density matrices using
the fidelity measure \cite{notefid}
\begin{equation}
  F[\Omega_m(\beta),\tilde{\rho}_m(\beta)] = \Tr\left(\sqrt\Omega_m(\beta)\tilde{\rho}_m(\beta)
\sqrt\Omega_m(\beta)\right)^{1/2} \;.
\end{equation}
Throughout this Section both $\Omega_n(\beta)$ and $\Omega_m(\beta)$
are set to the same temperature, since we want to identify when the
temperature is intensive.

We apply the above considerations to a 1-D spin chain characterized by
the Ising Hamiltonian in a transverse field with open boundary
conditions
\begin{equation}\label{Ising}
\hat{H}_n=\frac12\sum_{i=1}^{n-1}\sigma_x^i\otimes\sigma_x^{i+1}-\frac{h}{2}\sum_{i=1}^n\sigma_z^i,
\end{equation}
where $\sigma_i$ are the Pauli matrices, and $h$ gives the strength of
an external magnetic field. The system experiences a quantum -- \ie,
zero temperature -- phase transition when $h=h_c=1$~\cite{sachdev}. The
two-spin correlation functions are given by~\cite{barouch,nielsen}
\begin{align}
  \langle \sigma_x^i\sigma_x^{i+r}\rangle &=
  \begin{vmatrix} G_{-1} & G_{-2} & \cdots & G_{-r} & \\ G_0 & G_{-1}
    & \cdots & G_{-r+1} & \\ \vdots & \vdots & \ddots & \vdots & \\
    G_{r-2}& G_{r-3}& \cdots & G_{-1} &
  \end{vmatrix}, \label{eq:corr1} \\ \langle
  \sigma_y^i\sigma_y^{i+r}\rangle &= \begin{vmatrix} G_{1} & G_{0} &
    \cdots & G_{-r+2} & \\ G_2 & G_{1} & \cdots & G_{-r+3} & \\ \vdots
    & \vdots & \ddots & \vdots & \\ G_{r}& G_{r-1}& \cdots & G_{1} &
  \end{vmatrix},
  \label{eq:corr2} \\ \langle \sigma_z^i\sigma_z^{i+r}\rangle &=
  4\langle\sigma_z\rangle^2 - G_rG_{-r} \label{eq:corr3},
\end{align}
where
\begin{align} \nonumber G_r =& -\frac{1}{\pi}\int_0^\pi d\phi\,
  \cos(\phi r)
  (\cos\phi-h)\frac{\tanh(\frac{1}{2}\beta\omega_\phi)}{\omega_\phi}
  \\ \label{eq:corr4} &+\frac{1}{\pi}\int_0^\pi d\phi\,
  \sin(\phi
  r)\sin(\phi)\frac{\tanh(\frac{1}{2}\beta\omega_\phi)}{\omega_\phi},\\
  \omega_\phi =& \sqrt{(\sin \phi)^2+(h-\cos
    \phi)^2}.
\end{align}
The parameter $r$ sets the distance between the particles, \eg~$r=1$
means two neighbor particles. These correlators are calculated for
chains in the thermodynamic limit (\ie, $n\rightarrow \infty$). Using
these formulae, one can compute the reduced density matrix for the
2-spin system $\tilde{\rho}_2(\beta)$,
\begin{equation}
\tilde{\rho}_2(\beta) = \frac{1}{4} \left[\mathbb{I} + \sum_{i,j}
\langle\sigma_i^k \sigma_j^{k+1}\rangle \sigma_i \otimes
\sigma_j\right], \label{eq:rho}
\end{equation}
 without the explicit construction
of the global thermal state of $n$ particles \cite{nielsen}.

The minimal size for which both the terms in the Hamiltonian
(\ref{Ising}) contribute to the construction of $\Omega_m(\beta)$ is
for $m=2$. We then devote much attention to this first non-trivial
case of two-spin blocks. Furthermore, it is reasonable to expect that,
if for such small blocks the temperature is intensive, it will be
intensive for larger blocks as well. This intuition will be confirmed
in Section \ref{larger}, where the size dependence of our
considerations will be analyzed.

Before proceeding with the results, let us comment about our choice of the
local Hamiltonian $\hat{H}_2$, from which the reference state $\Omega_2(\beta)$
is derived. With this choice we are at first sight disregarding the interaction
between the $2$-spin block and the rest of the system. Whereas this may be
easily justified in the case of large $m$ and $n$, it deserves to be clarified
in our setting. In fact, we considered various strategies in order to
take into account the interactions at the border of the block,
in particular a mean-field approach, and the limit of high temperatures.
\par
First, one may add to $\hat{H}_2$ a correction term which takes into account the
surrounding spins of the block following a mean-field approach. The idea is to replace the operators at the boundaries by their mean
value $\langle\sigma_x\rangle$, so the corresponding effective two-particle Hamiltonian reads
\begin{eqnarray}
\hat{H}'_2&=&\frac12 \sigma_x\otimes\sigma_x - \frac{h}{2}\sum_{i=1}^2 \sigma_z^i + \frac12\sum_{i=1}^2\langle\sigma_x\rangle \sigma_x^i
\end{eqnarray}
The boundary correction term however turns
out be zero for any finite temperature, as $\langle\sigma_x\rangle=0$. This can be seen by considering that the
canonical distribution $\Omega_n(\beta)$ inherits the symmetries of the Ising
Hamiltonian, in particular the global spin-flip symmetry $U=\bigotimes_i \sigma_z^i$. As
a consequence, one has that $[U,\Omega_n(\beta)] = 0$, which implies that
\begin{equation}\label{sigmax}
[\sigma_z,\tilde{\rho}_1(\beta)] = 0
\end{equation}
for any finite $\beta$. Considering now that $\tilde{\rho}_1(\beta)$ can be expanded in
terms of the Pauli matrices, \refeq{sigmax} imposes that $\langle\sigma_x\rangle=0$ for any $T>0$ (see
also Ref.~\cite{nielsen}). However, recall that 
a symmetry breaking can occur at $T=0$, resulting in $\langle\sigma_x\rangle \neq0$.

Second, one could consider a correction valid for high temperatures.
A first order expansion for $\beta\rightarrow 0$
immediately reveals that in this limit $\tilde{\rho}_2(\beta)$ is given by
\begin{eqnarray}
\tilde{\rho}_2 (\beta)&\simeq& \Tr_{n-2}\left(\frac{\mathbb{I}^{\otimes n}}{2^n}-
\beta\hat{H}_n\right) \nonumber \\
 &=& \frac{\mathbb{I}^{\otimes2}}{2^2}- \beta\hat{H}_2
\end{eqnarray}
since the Pauli matrices appearing in the interacting terms of $\hat{H}_n$ are
traceless. The expression above for $\tilde{\rho}_2(\beta)$ coincides with the high-T
expansion of $\Omega_2(\beta)$. In other words, for high temperatures the standard
situation is retrieved, and no correction has to be taken into account.

In order to further clarify our choice of the local Hamiltonian, let us consider the
classical antiferromagnetic one-dimensional Ising model, given by:
\begin{equation}\label{Ising_classical}
  H_n^{\rm cl}=\sum_{i=1}^{n-1}s^i s^{i+1}\;,
\end{equation}
where $s^i=\pm 1$. Notice that the model above gives the classical limit
of $\hat{H}_n$ for the case $h=0$ \cite{note0}. It is worthwhile to briefly consider
this example since it gives a relevant classical model whose local Hamiltonian contains
 no correction due to the boundary terms.
To show this, let us consider the thermal state associated to Hamiltonian
(\ref{Ising_classical}), described by a thermal probability distribution:
\begin{equation}\label{thermal_classical}
  P(s^1,...,s^n)=\frac{1}{Z_n} \exp\left[-\beta H_n^{\rm cl}\right] \;,
\end{equation}
where $Z_n$ is the partition function. A straightforward calculation shows that,
by summing over all the possible configuration of the $n-2$ spins surrounding
an arbitrary $2$ spin block, one obtains the following distribution function:
\begin{equation}\label{thermal_classical_2}
  P(s^k,s^{k+1})=\frac{1}{Z_2} \exp\left[-\beta H_2^{\rm cl}\right] \;,
\end{equation}
where $k$ is arbitrary ($1<k<n$), $H_2^{\rm cl}=s^k s^{k+1}$ is the local Hamiltonian
and $Z_2$ its respective partition function. As said, we see that for this kind of
classical system no correction due to the boundary terms is expected. Furthermore,
notice that \refeq{thermal_classical_2} implies that the temperature is intensive.
However we should stress here that this result is not true for a generic classical
system. In conclusion, given the considerations above, it seems reasonable to consider
the Hamiltonian $\hat{H}_2$ -- without any additional correction -- to build the
reference state $\Omega_2(\beta)$. We will adopt this choice throughout the paper.
Finally, let us notice that the same considerations apply for any size $m$ of the block
(see Sec. \ref{larger}).

\begin{figure}
\includegraphics[width=8cm]{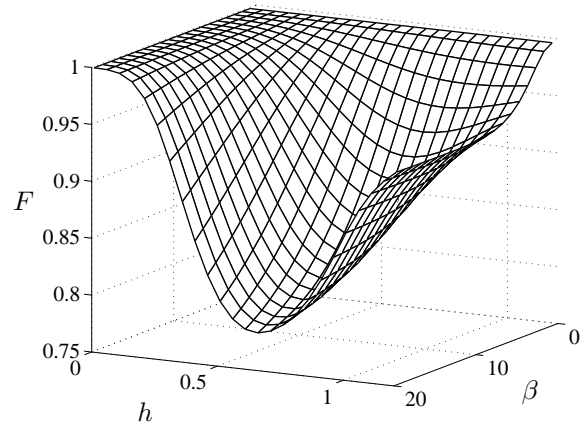}
\caption{\label{fig:bigfig} Fidelity
  $F[\Omega_2(\beta),\tilde{\rho}_2(\beta)]$ between the states
  $\Omega_2(\beta)$ (the thermal state of two spins at temperature
  $T=1/\beta$) and $\tilde{\rho}_2(\beta)$ (the two-spin reduced state
  of a global thermal system at temperature $T$) as a function of
  $\beta$ and the local magnetic field $h$ of the Ising Hamiltonian
  (\ref{Ising}). The temperature ceases to be an intensive magnitude
  when the fidelity is significantly different from one.}
\end{figure}
The results for the fidelity $F[\Omega_2(\beta),\tilde{\rho}_2(\beta)]$ are
plotted in Fig.~\ref{fig:bigfig}. We can see that it turns to be a non-trivial
function of the external magnetic field $h$ and the system global
temperature $T$. In particular we observe a high fidelity above a
certain temperature, for any value of $h$. This confirms the intuition
that for high temperatures the classical behavior should be recovered.
Namely, the reduced states are well approximated by thermal states at
temperature $T$. On the other hand, as we lower the temperature, the
fidelity can drop to values sensibly lower than one, thus indicating
that the standard thermodynamic description of the reduced state is no
longer accurate. We then recover the main feature of the results given
in \cite{malPRL}: the validity of the concept of intensive temperature
depends on the temperature itself, a behavior with no classical
analogue. Thanks to the sensitivity of the fidelity measure, we can
also analyze in details such a behavior, as can be seen for low
temperatures. In this case, the fidelity turns to be equal to one for
$h\gg1$ and $h\ll1$, as can be expected recalling that in both cases
the ground state is factorized (all the spins are aligned along the
same direction) \cite{sachdev}. On the other hand, for intermediate
magnetic fields, the fidelity drops to values sensibly lower than one,
showing a minimum which depends on the temperature.
We recall that the model in \refeq{Ising} has a critical point when $h_c=1$.



\begin{figure}
\includegraphics[width=8cm]{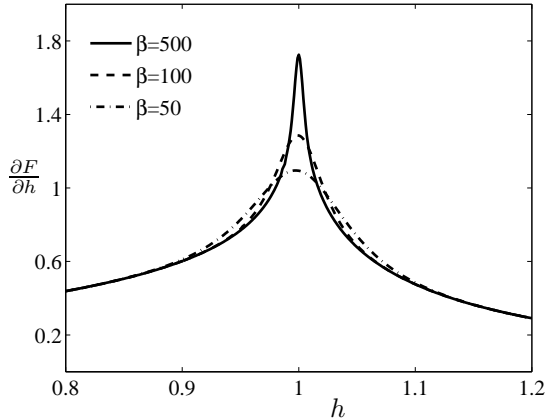}
\caption{\label{fig:fig5} First derivative of the fidelity
  $F[\Omega_2(\beta),\tilde{\rho}_2(\beta)]$ with respect to $h$ and
  for fixed values of $\beta$ ($\beta=50, 100, 500$). The behavior
  turns singular at the critical point $h_c$ for values of $\beta$
  closer to the ground state.}
\end{figure}

Before proceeding further, we now explicitly assess the sensitivity of
the fidelity in this scenario. For this purpose we numerically
evaluated the first derivative of the fidelity with respect to $h$ at
fixed $\beta$. We see in Fig.~\ref{fig:fig5} that, as $\beta$
increases, the maximum of the derivative gets more pronounced.
This behavior reflects the fact that the changes in the
ground state become sharper and sharper as the system
approaches criticality. As a consequence, subparts of the system
change sharply as well and the approximation of them with a thermal
state becomes more sensitive to small changes of $h$. In turn, the
actual value of the maximum in Fig.~\ref{fig:fig5} also increases as
the temperature decreases. This behavior proceeds until we reach some
$\beta$, which corresponds to some effective ground state, and below
which additional changes are hardly observed. This can be understood
by looking at the temperature dependence in the correlators in
\refeq{eq:corr4}. For large $\beta$, one can well approximate
$\tanh(\frac{1}{2}\beta\omega_\phi)\approx 1$, so the reduced states
become almost independent of the temperature. The derived magnitudes
computed from the reduced states exhibit for this reason minor changes
below some temperature.

\begin{figure}
\includegraphics[width=8cm]{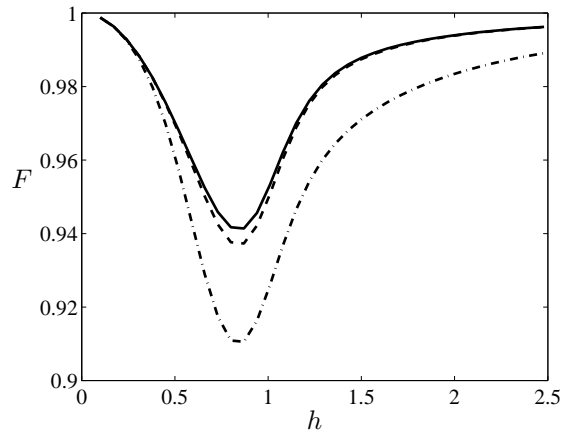}
\caption{\label{fig:sdistfid} For fixed $\beta=10$ we plot the
  fidelity $F[\Omega_{2,r}(\beta),\tilde{\rho}_{2,r}(\beta)]$ between
  the states $\Omega_{2,r}(\beta)$ [the reduced state of two particles
  distant $r$ obtained from a thermal state $\Omega_{r+1}(\beta)$] and
  $\tilde{\rho}_{2,r}(\beta)$ [the reduced state reduced state of two
  particles distant $r$ obtained from a thermal state
  $\Omega_{n}(\beta)$ with $n\rightarrow\infty$] for $r=3,5,7$ (bottom
  to top). Recall that $r$ denotes the distance between two spins. }
\end{figure}

The above considerations can be extended as well to two spins
separated by $r$ particles in the chain. In particular, we considered
a non-contiguous block of two distant spins which we denoted by
$\tilde{\rho}_{2,r}(\beta)$. We reconstruct the density matrix of such
a block using again \refeq{eq:rho} and considering the explicit
dependence on $r$ of the correlators [see \refeq{eq:corr4}].  Clearly,
it is no longer interesting now to compare $\tilde{\rho}_{2,r}(\beta)$
with $\Omega_ {2}(\beta)$. A much reasonable strategy is instead to
construct a thermal state $\Omega_ {r+1}(\beta)$ composed by $r+1$
spins and trace out all the particles but the two extremal ones. We
denoted the two-spin state so obtained as $\Omega_{2,r}(\beta)$. Then,
we can compare $\Omega_{2,r}(\beta)$ and $\tilde{\rho}_{2,r}(\beta)$
by calculating the fidelity
$F[\Omega_{2,r}(\beta),\tilde{\rho}_{2,r}(\beta)]$. The results for
different values of $r$ are shown in Fig.~\ref{fig:sdistfid}.
Clearly the fidelity is higher with respect to $r=1$.
However, some qualitative features already observed in the case
$r=1$ are still present for larger $r$. In particular, the fidelity turns
equal to one for $h\gg1$ and $h\ll1$,
whereas for intermediate magnetic fields it shows a minimum.

%

It is worth summarizing at this point the two main features disclosed
by the above analysis. Namely, {\em i)} the intensive nature of the
temperature depends on the global temperature of the system itself,
and {\em ii)} the temperature ceases to be intensive in a limited
region of the phase space, namely for intermadiate magnetic fields
around the zero-$T$ critical point. As we will
mention later on, we obtained similar results also in systems different
from the one given by \refeq{Ising}.

\section{Effective local temperature}\label{local}

We have seen in the preceding section that, for some values of the
Hamiltonian parameters and the temperature, a two-particle block of a
thermal state may be different from a two-particle thermal state under
the same Hamiltonian at the same temperature. Now, we focus on the
reduced state $\tilde{\rho}_m$ and look for a valid description of it
in terms of only a few thermodynamic magnitudes. A key point in the
attempt to describe the reduced states is the fact that, under some
natural circumstances, they will become with high probability close to
a thermal state \cite{psw,lebo}. This follows from the general
structure of the Hilbert space of the global system, and the
restriction over the total energy \cite{note1}. In this direction, the
results in Refs.~\cite{psw,lebo} are quite general but their
application to a specific system has to be taken cautiously.  In
particular, a thermal state is recovered when the interaction between
the parts of the system under consideration is negligible with respect
to the interactions inside the parts themselves. 
This condition, however, may not always be satisfied. When the parts taken
into account are composed of two spins only, as analyzed in the
previous section, it is then non trivial that the reduced states are
actually in a thermal state of their respective Hamiltonian.
Nonetheless, we will see that this is indeed the case in the majority
of the circumstances.

Following these considerations we check whether the reduced state is a
thermal state as well, \emph{but at an effective local temperature}
different from the one of the global system. That is, even if the
temperature may not be intensive, the reduced states may still have a
simple thermodynamical description. To this end, we consider a generic
$m$-particle thermal state $\Omega_m(\beta')$ and look for the
effective optimal temperature $\tilde\beta$ for which such a state
better describes the actual $m$-particle reduced state
$\tilde{\rho}_m(\beta)$. As outlined in the previous
Section, we assume that the local interaction
$\hat{H}_m$ is the same as in the global system $\hat{H}_n$ (actually
$\hat{H}_m$ is obtained from $\hat{H}_n$ by tracing out the $n-m$
disregarded spins), we only have to adjust the temperature.  In order
to check the quality of this description, we compute the fidelity
between the reduced state and the reference thermal state, and then
optimize the temperature of the latter. That is, once we compute the
traced state $\tilde{\rho}_m(\beta)$ for any given $\beta$, we
optimize over the parameter $\beta'$ the fidelity function
$F[\Omega_m(\beta'),\tilde{\rho}_m(\beta)]$. In this way, as said, we
identify an effective local temperature $\beta'=\tilde{\beta}$ for the
reduced state of the system.


\begin{figure}
\includegraphics[width=8cm]{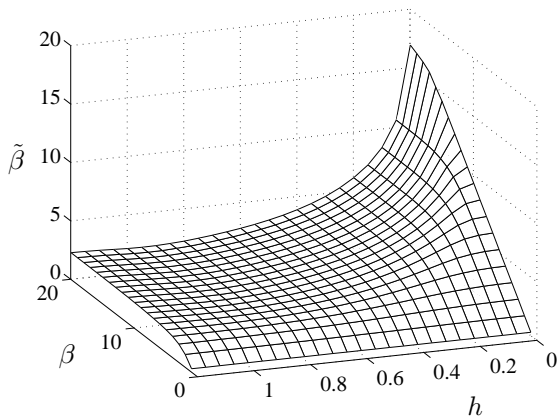}
\caption{\label{fig:fig8} The optimized inverse temperature
  $\tilde{\beta}$ (local temperature) of a reduced subsystem of two
  spins as a function of $\beta$ ($\beta=1/T$, where $T$ is the global
  temperature of the system), for different values of $h$.}
\end{figure}

Operating similarly as in the preceding section, we study for $m=2$
the relation between $\beta$ and $\tilde{\beta}$. As shown in
Fig.~\ref{fig:fig8}, we have that $\tilde{\beta}\approx\beta$ for low
values of $\beta$ (\ie, the local temperature is the same as the
global one). Thus, this range of temperatures can be identified as a
classical regime, where both the local and the global temperature
coincide, and the temperature behaves as an intensive magnitude. By
lowering the temperature this equality stops to hold, and the value of
$\tilde{\beta}$ saturates.  For even lower temperatures
of the global system, the reduced state keeps the same effective
temperature, even in conditions of zero-$T$ where the whole chain is
in its ground state. As above, this is due to the fact that for
sufficiently small $T$, the reduced states hardly change with
temperature. Notice also that the relation
  $\tilde{\beta}\approx\beta$ holds even for small temperature when the
  magnetic field vanishes (\ie, $h\ll1$). This is true also for
  $h\gg1$ (not shown in the picture) and can be understood in view of
  the considerations made in the previous Section.


\begin{figure}
\includegraphics[width=8cm]{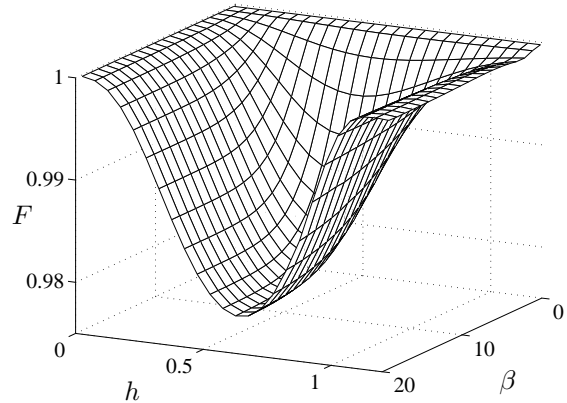}
\caption{\label{fig:fig6} Optimized fidelity $F_{\rm
    opt}=F[\Omega_2(\tilde{\beta}),\tilde{\rho}_2(\beta)]$ between the
  thermal state $\Omega_2(\tilde{\beta})$ at an optimized inverse
  temperature $\tilde{\beta}$ (local temperature) and the reduced
  state $\tilde{\rho}_2(\beta)$ of two spins at inverse temperature
  $\beta$ (global temperature). Compared with Fig.~\ref{fig:bigfig} the
  values of the fidelity are sensibly higher.}
\end{figure}

\begin{figure}
\includegraphics[width=8cm]{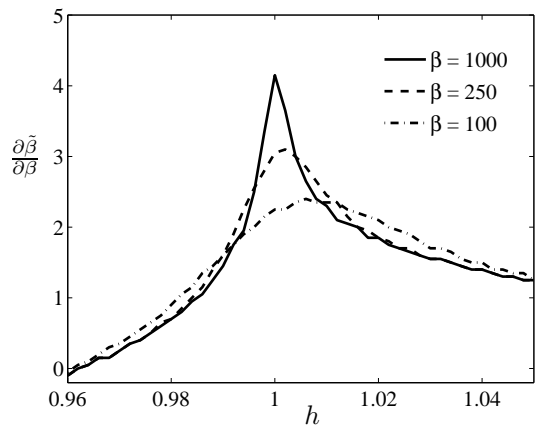}
\caption{\label{fig:fig4.eps} Value of $\frac{\partial
    \tilde{\beta}(\beta)}{\partial h}$ as a function of the local
  field $h$ ($\beta=100,250,1000$).}
\end{figure}

Performing the optimization over the local temperature we improved
dramatically the previous values of the fidelity, as can be seen
comparing Figs.~\ref{fig:bigfig} and \ref{fig:fig6} \cite{notefid2}.
As a matter of fact, after the local optimization the fidelity is
everywhere almost one, attaining the minimum of
$F[\Omega_2(\tilde{\beta}),\tilde{\rho}_2(\beta)]\approx0.975$ for
$h\approx0.6$. Such result implies that a local temperature
$\tilde{\beta}$ may be defined for almost every $\beta$ and $h$, even
if such local temperature is no longer intensive. In other words, the
findings in Refs.~\cite{psw,lebo} can be applied, at least
approximately, also in cases like the one studied here. This is in a
sense surprising, since here the assumptions used to derived the
results of Refs.~\cite{psw,lebo} are no longer satisfied. In
particular, the block we are considering is composed by only two
particles, implying that the interaction between it and the rest of
the system cannot be disregarded
a priori. Let us notice moreover that the definition of a local temperature
is not completely satisfactory for intermediate magnetic field $h$ at low
temperature. This resembles the results obtained in the previous Section even
if, as said, the fidelity here obtained is much higher.
We will commment on this in Sec.~\ref{esco}.


In order to check the sensitivity of the local temperature
$\tilde{\beta}$ to small changes of the parameters we studied the
derivative $\frac{\partial \tilde{\beta}(\beta)}{\partial h}$ as a
function of the local field $h$ which drives the quantum phase
transition.  We plot the results in Fig.~\ref{fig:fig4.eps}, where a
singular value around the critical point appears. This shows that
minor changes in the global temperature will strongly change the local
thermal properties of the reduced subsystem. The comparison of
Fig.~\ref{fig:fig5} and \ref{fig:fig4.eps} shows that the influence at
finite $T$ of the quantum phase transition is more pronounced for
local effective temperatures.

\begin{figure}
\includegraphics[width=8cm]{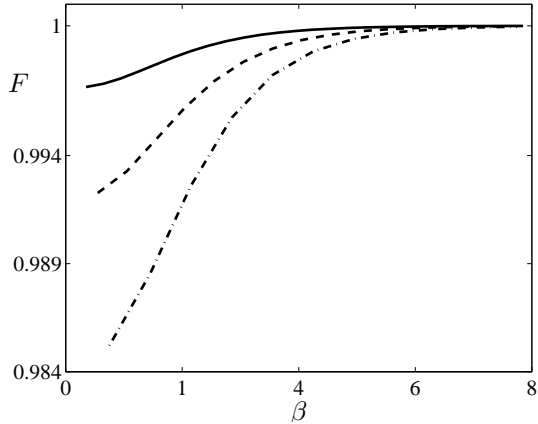}
\caption{\label{fig:fig7} Fidelity
  $F[\tilde{\rho}_2(\beta),\tilde{\rho}_2(\beta+\Delta \beta)]$
  between the two reduced states of a system at inverse temperature
  $\beta$ and $\beta+\Delta \beta$ respectively ($\Delta \beta = 0.3,
  0.5, 0.7$ from top to bottom). The magnetic field is set to
  $h=0.1$.}
\end{figure}

We complete the study of two-particle reduced states comparing two of
them at slightly different temperatures. In particular, we compute the
fidelity $F[\tilde{\rho}_2(\beta),\tilde{\rho}_2(\beta+\Delta \beta)]$
between two reduced states at temperatures $\beta$ and $\beta+\Delta
\beta$, as a function of the global system temperature $\beta$. Here
$\tilde{\rho}_2(\beta)$ is again a two-particle reduced state of a
chain at global temperature $\beta$. In Fig.~\ref{fig:fig7} we show
the fidelity as a function of $\beta$ for different values of $\Delta
\beta$.
The value of $\beta$ for which
$F[\tilde{\rho}_2(\beta),\tilde{\rho}_2(\beta+\Delta \beta)]=1$,
combined with the preceding results, suggest that above a given
$\beta$ the states are equal and independent of the temperature. This
is a direct indication of the asymptotic character of the reduced
states $\tilde{\rho}_2(\beta)$ for low temperatures, as already
expected from Eq.~(\ref{eq:corr4}).  In addition, the values of the
fidelity for different $\Delta \beta$ appear converging to one at a
value $\beta$ consistent with the saturation of the local temperature
found in this Section.

\section{Analysis for increasing block size}\label{larger}

In the previous Sections we mainly discussed reduced blocks
consisting of two spins, not necessarily contiguous. Now, we extend our considerations
to larger blocks. Recall that in the case of two-particle blocks
we considered the whole system in the thermodynamic limit
($n\rightarrow\infty$) and were able to construct the density matrix of
the block via the two-body correlators in
Eqs.~(\ref{eq:corr1})-(\ref{eq:corr3}). Even if such a procedure can
be in principle extended to higher order correlators (for the 3-spin
correlators see for instance \cite{patane}), it is more
convenient to use a different strategy when
blocks of arbitrary length are considered. Specifically, we use the formalism
of Matrix Product States (MPS), a numerical tool which has been shown to
describe with high precision ground and thermal states of 1D local
Hamiltonians. Using the MPS formalism for mixed states \cite{mpdo,vidal} we
construct thermal states and compute correlation functions in an
efficient way. Within the adopted numerical approach, we cannot
work directly in the thermodynamical limit.  However, the MPS
formalism allows to simulate systems large enough to accurately
reproduce the thermodynamical limit. We checked that systems composed
by around $50$ spins already suffice to obtain results
indistinguishable to the corresponding ones in the thermodynamical
limit. We will explicitly show in the following such a comparison.

The interest in considering blocks of increasing size is motivated
by the following observations. On the one hand, when the block
increases in size the interaction between the latter and the rest of
the system becomes less relevant when compared with the interactions
inside the block itself. This may intuitively lead to guess that
the rest of the system plays a negligible role in the thermalization
process. In other words, each block in which the global
system may be divided could be considered as a system on its own, thus
thermalizing with the global environment independently from the rest.
In such a scenario the temperature would clearly be intensive, for
large enough block size. On the other hand, there are
situations in which it is not possible to disregard any part of the
whole system, in particular when the latter is in a highly correlated
state (\eg, near a phase transition). Thus, an analysis of the
intensive nature of temperature as a function of the system size can
identify which of these two tendencies prevails and in which setting.

Before proceeding, let us briefly recall the MPS formalism (see
\cite{mpdo} for more details). This representation is based on a set
of matrices $A_k^s$ of size $D \times D$ used to write a quantum state
as
\begin{equation}
|\Psi_{MPS}\rangle = \sum_{s_1\ldots s_n=1}^d Tr(A_1^{s_1}\ldots A_n^{s_n})|s_1\ldots s_n\rangle.
\end{equation}
We can write mixed states in the MPS formalism introducing at each
position an ancillary system $a_k$ of dimension $d$. Thus, the thermal
state $\rho$ is written as a pure state in a Hilbert space of larger
dimension as
\begin{equation}
  |\rho\rangle = \sum_{s_1\ldots s_n}^d\sum_{a_1\ldots a_n}^d Tr(\prod_{k=1}^n A_k^{s_k,a_k})|s_1\ldots s_n\rangle|a_1\ldots a_n\rangle.
\end{equation}
Using this purification one can recover the thermal state tracing out the ancillary
systems $a_k$, $\rho = \Tr_a(|\rho\rangle\langle \rho |)$.

With this representation we can compute expectation values efficiently
using the relation
\begin{equation}
\langle O_1 \ldots O_n\rangle_\rho = Tr(E_{1,O_1}\ldots E_{n,O_n})\,,
\end{equation}
where $E_{k,O}=\sum_{s,s'}\langle s'|O|s\rangle M_k^{s,s'}$, and the
set of matrices $M_k$ are the result of tracing the ancilla states
\begin{equation}
M_k^{s,s'}=\sum_{a=1}^d A_k^{s,a} \otimes (A_k^{s',a})^*\,.
\end{equation}

With an initial MPS representation of a state at $\beta=0$, which
corresponds to the completely mixed state $\one$, we obtain the state
$\rho \propto \exp(-\beta H)$ using
\begin{equation}
e^{-\beta H} = (e^{-\Delta t H})^M \one (e^{-\Delta t H})^M,
\end{equation}
where $\Delta t = \beta / 2M$. In this way the thermal state at
temperature $\beta$ is the result of the evolution in imaginary time
of the completely mixed state, an evolution that can also be performed
in an efficient way by means of a Trotter decomposition of the time
evolution operator. Using this numerical technique we can extend the
calculation of the correlation functions up to $m=6$, the
computational limit being the construction of the state $\rho$ from
the corresponding correlators as in \refeq{eq:rho}.

\begin{figure}
\includegraphics[width=8cm]{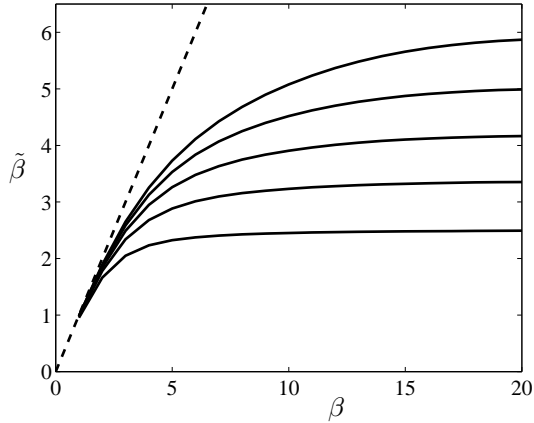}
\caption{\label{fig:fig2}The local inverse temperature $\tilde{\beta}$
  of a reduced subsystem of $m$ spins as a function of the global
  inverse temperature $\beta$ of the system (the magnetic field is set
  to $h = 0.8$; the matrices size for the MPS algorithm is set to
  $D=15$).  From bottom to top we set $m=2,3,4,5,6$. The dashed-dotted
  line shows $\beta=\tilde{\beta}$ for reference. We see that for
  larger blocks the local temperature gets closer to the global one.
  The actual region in which the temperature is intensive
  does not seem to be affected by the size of the block ($m>2$).}
\end{figure}

\begin{figure}
\includegraphics[width=8cm]{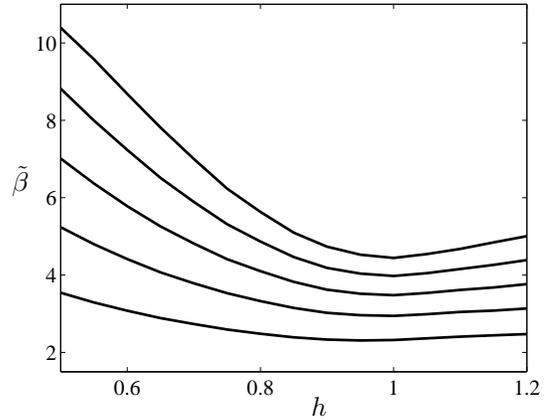}
\caption{\label{fig:fig3} We plot the local inverse temperature
  $\tilde{\beta}$ reached by the reduced subsystem of $m$ spins as a
  function of the magnetic field $h$ (the global inverse temperature
  is set to $\beta=15$). From bottom to top the size of the block
  increases ($m=2,3,4,5,6$).}
\end{figure}

We plot in Fig.~\ref{fig:fig2} the effective local temperature,
$\tilde{\beta}$, as a function of global one, $\beta$, for reduced
systems of size $m=2 - 6$. The first feature that one may note is that
the local optimized temperature tends to get closer to the global one
as the block size increases.  This is in agreement with the trivial
fact that for large block size the local and global temperature
coincide, as the block itself coincides with the whole system.
Furthermore, as mentioned above, as the block size increases the
interaction between the latter and the rest of the system becomes less
relevant. Thus, the block will be in a thermal state with high
probability, as a consequence of the results exposed in
Refs.~\cite{psw,lebo}. For each curve depicted in Fig.~\ref{fig:fig2},
we checked explicitly the fidelity between a thermal state at
temperature $\tilde{\beta}$ and the actual reduced state. We obtained
that, all along these curves, the fidelity is very close to one,
confirming that the results in Refs.~\cite{psw,lebo} can be applied to
the finite systems considered here.  In turn, this points to the fact
that the systems are already near to the thermodynamic limit, at least
for what concerns the properties of their reduced states. However, the
fact that the blocks are thermal does not mean that the temperature is
intensive. As said, Fig.~\ref{fig:fig2} shows that the local
temperature get closer to the global one for larger blocks,
nevertheless the actual region in which the temperature is intensive
does not seem to be much affected by the size of the blocks.
This suggests that the quantum fidelity analysis reveals
fine features of the systems that cannot be understood by
considerations regarding only the energy balance between the various
parts of the system. In particular, as mentioned above, correlations
should be considered in order to clarify such a behavior.

We also notice that for each value of $m$ a different saturation value
of $\tilde{\beta}$ is obtained. The latter corresponds to the
effective temperature of the reduced states when the whole system is
at zero $T$. Thus for the ground states of the considered systems, the
temperature is clearly not intensive, as expected. On the other hand,
for high enough temperature, we can identify again a classical regime
--- \ie, a regime in which the temperature is intensive. In general,
outside this classical region we observe that the local temperature is
sensibly higher than the global one.

Before concluding this Section let us consider the influence of the
zero-$T$ properties of the system at small temperatures. As seen in
the previous Section (see Fig.~\ref{fig:fig4.eps}), we expect that at
small temperatures the features of the system ground state can be
revealed by an analysis of the local temperature as a function of the
local field $h$. The results for $\beta = 15$ are plotted in
Fig.~\ref{fig:fig3}. In accordance with the results reported in
Fig.~\ref{fig:fig2}, as the block size increases the local temperature
gets closer to the global ones for any value of $h$. Furthermore,
notice that, as the block size increases, the minimum of
$\tilde{\beta}$ (that identifies when the temperature deviates mostly
from an intensive behavior) get closer to $h_c$. This behavior around
the critical value of the field $h=1$ suggests that zero-$T$
properties of the system ground state have an important influence on
the thermal state of the block.

\section{Conclusions}\label{esco}

With the study of many-body systems using thermodynamic quantities,
one may obtain global properties of a system by measurements performed
only on a local (reduced) part of it.  However, depending on the
conditions of the physical system under consideration, some of the
fundamental assumptions of statistical equilibrium may fail and a
valid thermodynamical description cannot be possible. We have
identified this situation in spin chains at low temperatures, where
the reduced states are not thermal states at the same temperature as
the global system. In particular, we have seen that the temperature
ceases to be an intensive magnitude in dependence of the global
temperature itself, a feature without classical analogue.  This result
is in accordance with the findings reported in Ref.~\cite{malPRL}.
Notice, however, that here we made no assumption on the correlations
between the blocks in which the global system may be decomposed. In
fact, we focused only on the reduced local state of the system. Our
approach is motivated by an operational viewpoint, since an actual
measure performed on a local part of a system gives no information
about correlations with other parts of it.

Remarkably, we have seen that an effective thermodynamical description
of the reduced states is often still possible, despite the temperature
may not be an intensive magnitude. In particular, the resulting reduced
states can be described with high accuracy as thermal states at an optimized
temperature. Interestingly, this is valid \textit{i)} without any additional
assumption on the interactions, and \textit{ii)} even if we considered
blocks of particularly small size (\ie, composed of $2$ to $6$ spins).
This local temperature becomes of course equal to the global
one in the regime of high temperatures, where we recover the
classical behavior. However, by lowering the temperature of the global
system, the local temperature saturates at some point. In other words,
depending on the system parameters, the local temperature may be
different from zero even when the global system is in the ground
state. The dependence of the local temperature on the size of the blocks has
been studied too. As one may expect, the local temperature get closer
to the global one for larger blocks.

We pointed out that the departure from the classical behavior is more
pronounced at low temperatures and for intermediate values of the local
magnetic field, around the zero-$T$ critical point (this is true concerning
both the intensive behavior of temperature and the definition of a local
temperature, as can be seen in Figs.~\ref{fig:bigfig}, \ref{fig:sdistfid},
and \ref{fig:fig6}). Notice that this is the region of parameters where
quantum correlations are expected to be stronger. For example, at zero temperature
the pair-wise entanglement between two spins shows its higher values in
this region \cite{qpt,nielsen}, as well as the entanglement between a block
of spins and the rest \cite{latorre}.
Though the existence of a finite block entropy may suggest that entanglement
plays a crucial role in all these results, preliminary studies indicate that
the relation between quantum correlations and the definition of a local temperature
is nontrivial.
Thus it will be a subject of further studies the way in which quantum and classical
correlations lead to the bahaviours identified here.

Let us mention here that we have analyzed also other Hamiltonian systems.
In this paper we have shown a study of the transverse Ising model, but the extension
to a generic XY interaction leads to similar results and conclusions.
Applying the MPS techniques we can extend the analysis to spin models with
arbitrary interaction, such as spin chains with Heisenberg interaction.
Furthermore, we analyzed chains consisting of harmonic oscillators with quadratic
interaction (harmonic chains). In all these cases,
the obtained results were qualitatively very similar to those shown here.

In order to assess the canonical character of the reduced states we
employed the quantum fidelity, a quantity extensively used in quantum
information science. We have seen that the fidelity is particularly
suitable when studying the limits of applicability of the concept of
intensive temperature. In particular, harnessing the sensitivity of
quantum fidelity, we have observed how the low temperature features
of the systems influence our results.
This suggests that the quantum fidelity analysis reveals
fine features of the systems that cannot be understood by
considerations regarding only the energy balance between
the various parts of the system. In particular, as said, future investigations
should consider the role of correlations in order to clarify such a
behavior.

\acknowledgments We thank Marcelo Terra Cunha and Guifr\'e Vidal for
discussion. We gratefully acknowledge the financial support from the
EU Project QAP, Spanish MEC projects FIS2007-60182 and Consolider
Ingenio 2010 ``QOIT'', the ``Juan de la Cierva'' grant, the Grup
Consolidat de Recerca de la Generalitat de Catalunya and Caixa
Manresa.


\begin{thebibliography}{99}

\bibitem{MahlerBook} For an historical introduction see J. Gemmer, M.
  Michel, and G. Mahler, \emph{Quantum Thermodynamics} (Spinger, New
  York, 2004), and references therein.

\bibitem{tasaki} H. Tasaki, Phys. Rev. Lett. \textbf{80}, 1373 (1998).

\bibitem{malPRL} M. Hartmann, G. Mahler and O. Hess, Phys. Rev. Lett. \textbf{93}, 080402 (2004).

\bibitem{malPRE} M. Hartmann, G. Mahler and O. Hess, Phys. Rev. E \textbf{70}, 066148 (2004).

\bibitem{psw} S. Popescu, A. J. Short and A. Winter, Nature Physics \textbf{2}, 754 (2006).

\bibitem{lebo} S. Goldstein, J. L. Lebowitz, R. Tumulka and N. Zangh\`\i, Phys. Rev. Lett. \textbf{96}, 050403 (2006).

\bibitem{nanotherm} Y. Gao and Y. Bandao, Nature (London) \textbf{415}, 599 (2002).

\bibitem{mahlEPL} M. Hartmann and G. Mahler, Europhys. Lett. \textbf{70}, 579 (2005).

\bibitem{HartCP} M. Hartmann, Contemporary Physics \textbf{47}, 89 (2006).

\bibitem{NC} M. A. Nielsen and I. L. Chuang, \emph{Quantum Computation
    and Quantum Information} (Cambridge University Press, Cambridge,
  UK, 2000).

\bibitem{zan1} P. Zanardi, H. T. Quan, X. Wang and C. P. Sun, Phys. Rev. A \textbf{75}, 032109 (2007).

\bibitem{zan2} P. Zanardi and N. Paunkovic, Phys. Rev. E \textbf{74}, 031123 (2006).

\bibitem{olav} S. O. Skr\o vseth, Europhys. Lett. \textbf{76}, 1179 (2006).

\bibitem{HLW06} P. Hayden, D.W. Leung, and A. Winter, Commun. Math.
  Phys. \textbf{265}, 95 (2006).

\bibitem{schro} E. Schr\"odinger, \emph{Statistichal Thermodynamics},
  Dover Publications (1989).

\bibitem{notefid} In this work, we take as measure of distance between two
  quantum states the Uhlmann fidelity, defined as $F_U(\rho_1,\rho_2)=\Tr\left(\sqrt\rho_1\rho_2\sqrt\rho_1\right)^{1/2}$. We stress that other measurements of proximity between states, such as the fidelity functions derived from the trace distance, $F_{\rm Tr}(\rho_1,\rho_2) = 1 - \frac12\Tr[|\rho_1 -  \rho_2|]$, or the Hilbert-Shmidt distance $F_{\rm HS}(\rho_1,\rho_2) = 1 - \frac12\Tr[(\rho_1 -  \rho_2)^2]$, or the relative entropy $S(\rho_1||\rho_2)=\Tr[\rho_1(\log \rho_1-\log \rho_2)]$, lead to the same main results.

\bibitem{sachdev} S. Sachdev, \emph{Quantum Phase Transitions}
  (Cambridge University Press, Cambridge, 1999).

\bibitem{barouch} E. Barouch, B. McCoy and M. Dresden, Phys. Rev. A \textbf{2}, 1075 (1970);
E. Barouch and B. McCoy, Phys. Rev. A \textbf{3}, 786 (1971).

\bibitem{nielsen} T.J. Osborne and M.A. Nielsen, Phys. Rev. A \textbf{66}, 032110 (2002).

\bibitem{note0} We are omitting the local terms since we focus on possible corrections due to interactions at the boundary of the considered block. In this case (\ie, $h=0$) the model presented in \refeq{Ising_classical} gives the classical limit of the quantum model. The classical limit of the full quantum Hamiltonian $\hat{H}_n$ can be found in A. Cuccoli, A. Taiti, R. Vaia, and P. Verrucchi, Phys. Rev. B \textbf{76}, 064405 (2007).
However, the analysis of the latter goes beyond the scope of the present contribution.


\bibitem{note1} Here we are not considering systems for which an exchange
  of particles is possible between subparts of them or between the system
  and its environment.

\bibitem{notefid2} Clearly, such an improvement is of relevance only in
  those cases where the fidelity
  $F[\Omega_2(\beta),\tilde{\rho}_2(\beta)]$ is significantly smaller
  than one. In particular, our calculations show an improvement in the
  fidelity also for $h\gg1$. However, in these cases,
  $F[\Omega_2(\beta),\tilde{\rho}_2(\beta)]$ is already almost
  indistinguishable from one. The improvement obtained optimizing the
  local temperature is then insignificant.

\bibitem{patane} D. Patan\'e , R. Fazio and L. Amico,
 New J. Phys. \textbf{9}, 322 (2007).

\bibitem{mpdo} F. Verstraete, J. J. Garc\'\i a-Ripoll and J. I. Cirac,
  Phys. Rev. Lett \textbf{93}, 207204 (2004).

\bibitem{vidal}
M. Zwolak and G. Vidal, Phys. Rev. Lett \textbf{93}, 207205 (2004).

\bibitem{qpt} A. Osterloh, L. Amico, G. Falci and R. Fazio, Nature
  \textbf{416}, 608 (2002).

\bibitem{latorre} J. I. Latorre, E. Rico, and G. Vidal, Quant.~Inf.~Comput. \textbf{4}, 48 (2004).

\end{thebibliography}
\end{document}